\documentclass[aps, prl, twocolumn, floatfix, nofootinbib, superscriptaddress, longbibliography, amsmath, amssymb]{revtex4-1}

\usepackage{mathrsfs}
\usepackage{graphicx}
\usepackage{epsfig}
\usepackage{bm}
\usepackage{bbold}
\usepackage{color}
\usepackage{float}
\usepackage{dcolumn}
\usepackage{multirow} 
\usepackage{hyperref}
\usepackage{color}
\usepackage{braket}
\usepackage{dcolumn}
\usepackage{multirow}
\usepackage[titletoc]{appendix}
\usepackage{cleveref}
\usepackage{url}

\newcolumntype{d}[1]{D{.}{.}{#1}}

\begin{document}

\renewcommand{\arraystretch}{1.25}

\title{Two-neutrino double electron capture on \texorpdfstring{$\boldsymbol{{}^{124}}$}{}Xe\\
based on an effective theory and the nuclear shell model}

\author{E.\ A.\ Coello P\'erez}
\email{tcoello@theorie.ikp.physik.tu-darmstadt.de} 
\affiliation{Institut f\"ur Kernphysik, Technische Universit\"at Darmstadt, 64289 Darmstadt, Germany}
\affiliation{ExtreMe Matter Institute EMMI, Helmholtzzentrum f\"ur Schwerionenforschung GmbH, 64291 Darmstadt, Germany}
	
\author{J.\ Men\'endez}
\email{menendez@cns.s.u-tokyo.ac.jp}
\affiliation{Center for Nuclear Study, The University of Tokyo, Tokyo 113-0033, Japan}
	
\author{A.\ Schwenk}
\email{schwenk@physik.tu-darmstadt.de} 
\affiliation{Institut f\"ur Kernphysik, Technische Universit\"at Darmstadt, 64289 Darmstadt, Germany}
\affiliation{ExtreMe Matter Institute EMMI, Helmholtzzentrum f\"ur Schwerionenforschung GmbH, 64291 Darmstadt, Germany}
\affiliation{Max-Planck-Institut f\"ur Kernphysik, Saupfercheckweg 1, 69117 Heidelberg, Germany}

\begin{abstract}
We study the two-neutrino double electron capture on $^{124}$Xe based on an effective
theory (ET) and large-scale shell model calculations, two modern nuclear structure 
approaches that have been tested against Gamow-Teller and double-beta decay data.
In the ET, the low-energy constants are fit to electron capture and $\beta^-$ transitions 
around xenon. For the nuclear shell model, we use an interaction in a
large configuration space that reproduces the spectroscopy of nuclei in this mass region. For
the dominant transition to the $^{124}$Te ground state, we find half-lives 
$T^{2\nu {\rm ECEC}}_{1/2}=(1.3-18)\times 10^{22}\,$y for the ET and
$T^{2\nu {\rm ECEC}}_{1/2}=(0.43-2.9)\times 10^{22}\,$y for the shell model.
The ET uncertainty leads to a half-life almost entirely consistent with present experimental
limits and largely within the reach of ongoing experiments. The shell model half-life range
overlaps with the ET, but extends less beyond current limits.
Our findings thus suggest that the two-neutrino double electron capture on $^{124}$Xe
has a good chance to be discovered by ongoing or future experiments. In addition, we 
present results for the two-neutrino double electron capture to excited states of $^{124}$Te.
\end{abstract}

\maketitle

{\it Introduction.---}Second-order weak processes give rise to extremely rare decay 
modes of atomic nuclei. They have been observed in about a dozen nuclei with the
longest half-lives in the nuclear chart of about $10^{19}-10^{21}$ years~\cite{barabash2015}.
All these are two-neutrino double beta ($\beta\beta$) decays, where the emission
of two electrons is accompanied by two antineutrinos. An even rarer decay can occur
if no neutrinos are emitted, the neutrinoless $\beta\beta$ decay. This process is 
particularly intriguing, because neutrinoless $\beta\beta$ decay is not allowed in
the Standard Model, does not conserve lepton number, and can only happen if 
neutrinos are their own antiparticles (Majorana particles)~\cite{avignone2008}.

Due to its unique potential for neutrino physics, beyond the Standard Model physics,
and the understanding of the matter-antimatter asymmetry of the Universe, 
neutrinoless $\beta\beta$ decay searches are increasingly active~\cite{KamLAND-Zen16,EXO18,CUORE18,MAJORANA18,GERDA18,NEXT18,CUPID18}.
The planning and interpretation of these experiments relies on a good understanding
of the decay half-life, which depends on a nuclear matrix element. However, these
are poorly known as neutrinoless $\beta\beta$ decay matrix-element calculations
disagree by at least a factor two~\cite{engel2017}.

Second-order weak processes with neutrino emission are ideal tests of neutrinoless 
$\beta\beta$ decay matrix-element calculations. The initial and final states are 
common, and the transition operator is also very similar, dominated by the physics
of spin and isospin. In addition to $\beta\beta$ decay, a related mode is the 
two-neutrino double electron capture ($2\nu$ECEC). Here, two K- or L-shell
orbital electrons are simultaneously captured, rather than $\beta$ emitted. 
This mode is kinematically unfavored with respect to $\beta\beta$ decay, and 
at present only a geochemical measurement of $^{130}$Ba~\cite{Meshik2001, Pujol2009}, 
and a possible detection in $^{78}$Kr~\cite{Gavrilyuk2013,ratkevich2017} have 
been claimed. Moreover, a resonant neutrinoless ECEC could be fulfilled in selected nuclei~\cite{krivoruchenko2011,eliseev2011-1,eliseev2011-2}. For both ECEC modes
limits of $10^{21}-10^{22}$ years have been set in various isotopes~\cite{Angloher2016,Lehnert2016,agostini2016,Meshik2001,Barabash2011,Belli2013,Belli2016}.

$^{124}$Xe is one of the most promising isotopes to observe $2\nu$ECEC
due to its largest $Q$-value of $2857$~keV~\cite{Nesterenko2012}.
Large-volume liquid-xenon experiments primarily designed for
the direct detection of dark matter such as $\rm XMASS$~\cite{abe2013},
$\rm XENON100$~\cite{aprile2012}, or $\rm LUX$~\cite{akerib2014}
are sensitive to ECEC and $\beta\beta$ decays in $^{124}$Xe, $^{126}$Xe 
and $^{134}$Xe~\cite{mei2014, barros2014}. Enriched xenon gas detectors 
are also very competitive~\cite{Gavrilyuk2017,Gavrilyuk2018}. Recent searches
have reached a sensitivity comparable to the half-lives expected by most
theoretical calculations~\cite{abe2016,aprile2017}. Moreover the latest limits
set by the XMASS collaboration~\cite{abe2018} exclude most theoretical predictions.

Calculations of $2\nu$ECEC and two-neutrino $\beta\beta$ decay are challenging
because they involve the quantum many-body problem of heavy nuclei with 
even-even and odd-odd numbers of neutrons and protons.
The methods of choice are
the quasiparticle random-phase approximation (QRPA)~\cite{suhonen2012,simkovic2013},
extensively used for $^{124}$Xe $2\nu$ECEC~\cite{hirsch1994,suhonen2013,pirinen2015},
and the large-scale nuclear shell model~\cite{caurier2004}, which predicted 
successfully the $^{48}$Ca $\beta\beta$ half-live before its 
measurement~\cite{caurier1990,Poves1995}. In this mass region,
shell model studies have covered $^{124}$Sn, $^{128,130}$Te and
$^{136}$Xe~\cite{caurier2012,horoi2013,neacsu2015,horoi2016,coraggio2017},
but no shell-model calculation exists for $^{124}$Xe $2\nu$ECEC. In addition 
to QRPA, other more schematic approaches have also been applied to 
$^{124}$Xe~\cite{shukla2007, singh2007-1,rumyantsev1998,aunola1996}.

Further state-of-the-art $^{124}$Xe $2\nu$ECEC calculations are thus required 
given the tension between theoretical predictions and experimental limits.
In this Letter, we calculate the corresponding nuclear matrix elements
using an effective theory (ET), introduced in Ref.~\cite{coelloperez2017}, 
which describes well Gamow-Teller and two-neutrino $\beta\beta$
decays of heavy nuclei, including $^{128,130}$Te.
One of the advantages of the ET is to provide
consistent theoretical uncertainties.
Similar ETs have been used to study electromagnetic transitions in spherical~\cite{coelloperez2015-2, coelloperez2016} and deformed~\cite{papenbrock2011, zhang2013, papenbrock2014, coelloperez2015-1, papenbrock2015, chen2017, chen2018} nuclei.
In addition, we present the first large-scale nuclear shell model calculation
for $^{124}$Xe $2\nu$ECEC.
We focus on transitions to the $^{124}$Te $0^+_{\rm gs}$ ground state,
but also consider $2\nu$ECEC to the lowest excited $0^+_2$ and $2^+_1$ states.
The relation between the calculated nuclear matrix element $M^{2\nu{\rm ECEC}}$
and the $2\nu$ECEC half-life is given by
\begin{equation}
\big(T^{2\nu{\rm ECEC}}_{1/2}\big)^{-1} =
G^{2\nu {\rm ECEC}} g_A^4
\left|M^{2\nu{\rm ECEC}}\right|^2\,,
\end{equation}
where $G^{2\nu{\rm ECEC}}$ is a known phase-space factor~\cite{kotila2013}
and $g_A=1.27$ the axial-vector coupling constant.

{\it Effective theory.---}We use an ET that describes the initial $^{124}$Xe and
final $^{124}$Te nuclei, both with even number of protons and neutrons,
as spherical collective cores. The intermediate nucleus, $^{124}$I, has odd 
number of protons and neutrons. The ET describes its lowest $1^+_1$ state
as a double-fermion excitation of a $0^+$ reference state that represents the
ground state of either $^{124}$Xe or $^{124}$Te, $|1^+_1; j_p;j_n\rangle=
\left(n^\dagger\otimes p^\dagger\right)^{(1)} |0^+\rangle$. Depending on the
reference state, $n^\dagger$ ($p^\dagger$) creates a neutron (proton)
particle or hole in the single-particle orbital $j_n$ ($j_p$). At leading order,
higher $1^+$ states are described as multiphonon excitations,
with energies with respect to the reference state
$E(1^+_{n+1})=E(1^+_1)+n\omega$, where
$\omega$ is the excitation energy and $n$ is the number
of phonon excitations.

The effective spin-isospin ($\bm \sigma \bm \tau$) Gamow-Teller operator
is systematically constructed as the most general rank-one operator. At
leading order it takes the form~\cite{coelloperez2017}
\begin{align}
O_{\rm GT} &= C_{\beta}
\left(\tilde{p} \otimes \tilde{n} \right)^{(1)} \nonumber \\
& + \sum\limits_{\ell} C_{\beta \ell} \left[
\left(d^{\dagger} + \tilde{d}\right) \otimes
\left( \tilde{p} \otimes \tilde{n} \right)^{(\ell)} \right]^{(1)}
\nonumber \\
& + \sum\limits_{L\ell} C_{\beta L\ell} \left[
\left(d^{\dagger} \otimes d^{\dagger} +
\tilde{d}\otimes\tilde{d}\right)^{(L)} \otimes
\left( \tilde{p} \otimes \tilde{n} \right)^{(\ell)} \right]^{(1)}
\nonumber \\
& + \ldots,
\label{o_gt}
\end{align}
where the tilde denotes well defined annihilation tensor operators,
and the phonon ($\tilde{d}$, $d^{\dagger}$) and nucleon operators are
tensor coupled. The low-energy constants $C$ must be fitted to data, and
the expansion above is truncated after terms involving more than two phonon
creation or annihilation operators. The first term in Eq.~(\ref{o_gt}) couples the
reference state to the lowest $1^+_1$ state of the odd-odd nucleus, so that
$C_\beta$ can be extracted from the
known $\log(ft)$ value of the corresponding $\beta$ decay or EC:
\begin{equation}
\langle 0^+ |O_{\rm GT} | 1^+_1 \rangle =
\sqrt{\frac{3 \kappa}{g_A^2 10^{\log(ft)}}}\,,
\label{betaEC}
\end{equation}
where $\kappa=6147$ is a constant. The
power counting of the ET~\cite{coelloperez2015-2, coelloperez2016,
coelloperez2017} relates the Gamow-Teller matrix elements
from the lowest and higher $1^+$ initial states to the common final reference state by
$\langle 0^+ |O_{\rm GT} | 1^+_{n+1} \rangle \sim (\omega/
\Lambda)^{n/2} \langle 0^+ | O_{\rm GT} | 1^+_1 \rangle$,
where $\Lambda\sim3\omega$ is the breakdown
scale of the ET. This allows us to estimate the values of $C_{\beta\ell}$
and $C_{L\beta\ell}$ with consistent theoretical uncertainties.

The $2\nu$ECEC matrix element from the ground state of the initial nucleus
to a $0^+$ state of the final one is
\begin{align}
M^{2\nu{\rm ECEC}}=
&\sum_j\frac{\langle 0^+_{\rm{f}} | O_{\rm GT} | 1^+_j \rangle
\langle 1^+_j | O_{\rm GT} | 0^+_{\rm{gs,i}}\rangle}{D(1^+_j)/m_e} \,,
\label{M2EC_exact}
\end{align}
where $j$ sums over all $1^+_j$ states of the intermediate nucleus.
The electron mass $m_e$ keeps the matrix element dimensionless,
and the energy denominator is
$D(1^+_j)=E(1^+_j)-E(0^+_{\rm{gs,i}})+[E(0^+_{\rm{f}})-E(0^+_{\rm{gs,i}})]/2$,
neglecting the difference in electron binding energies.
The expression for the $2\nu$ECEC to a final $2^+$ state is similar~\cite{pirinen2015},
but the energy denominator appears to the third power.

Because the ET is designed to reproduce low-energy states,
we calculate the $2\nu$ECEC
matrix elements within the single-state dominance (SSD) approximation:
\begin{align}
M^{2\nu{\rm ECEC}}
\approx
\frac{\langle 0^+_{\rm{f}} | O_{\rm GT} | 1^+_1 \rangle
\langle 1^+_1 | O^+_{\rm GT} | 0^+_{\rm{gs,i}} \rangle}
{D(1^+_1)/m_e}\,,
\label{M2EC}
\end{align}
which implies that
only the matrix elements involving the lowest $1^+_1$ state contribute.
The advantage is that the ET can fit these using Eq.~(\ref{betaEC}).
The contribution due to omitted higher
intermediate $1^+$ states is estimated within the ET and
treated as a theoretical uncertainty~\cite{coelloperez2017}:
\begin{equation}
\frac{\Delta M^{2\nu{\rm ECEC}}}
{M^{2\nu{\rm ECEC}}} =
\frac{D(1^+_1)}{\Lambda}
\Phi \left(\frac{\omega}{\Lambda},1,
\frac{D(1^+_1)+\omega}{\omega}\right)\,,
\label{SSD_relative}
\end{equation}
where $\Phi(z,s,a)=\sum_{n=0}^\infty z^n/(a+n)^s$
is the Lerch transcendent. The ET describes very well the experimentally
known two-neutrino $\beta\beta$ decay half-lives once the ET
uncertainties, including from Eq.~(\ref{SSD_relative}), are taken into
account~\cite{coelloperez2017}. This agreement includes $^{128,130}$Te
among other heavy nuclei.

The ET $2\nu{\rm ECEC}$ matrix element calculation thus requires the 
known ground-state energies and the lowest $1^+_1$ excitation energy
to calculate the energy denominator, as well as the Gamow-Teller $\beta$-decay and EC
matrix elements from the $1^+_1$ to the initial and final states of the 
$2\nu$ECEC to fit the low-energy constants. In addition, the collective mode
$\omega$ sets the ET uncertainty.  Unfortunately, for $^{124}$Xe
there are no direct measurements for Gamow-Teller $\beta$ decay or
EC from the lowest $1^+_1$ state in $^{124}$I (the ground state is $2^-$),
or alternatively zero-angle charge-exchange reaction cross sections
involving the nuclei of interest. The $1^+_1$ excitation energy in $^{124}$I
is also unknown.

Therefore, we adopt the following strategy. First, we set 
$\log(ft)^{\rm EC}=5.00(10)$ for the EC on the lowest $1^+_1$ state in $^{124}$I,
based on the experimental range of known EC on iodine isotopes with nucleon 
number $A=122-128$. This quantity varies smoothly for nuclei within an isotopic
chain. For the $\beta^-$ decay, we set  $\log(ft)^{\beta^-}=1.06(1)\log(ft)^{\rm EC}$,
based on the systematics of odd-odd nuclei in this region of the nuclear chart.
Guided by the known spin-unassigned excited states
of $^{124}$I and the systematics in neighboring odd-odd
nuclei, we set the excitation energy of the first $1^+_1$
state in $^{124}$I as $105-170$~keV. Because this range is much smaller than
the energy differences in $D(1^+_1)$, the associated uncertainty in the matrix elements
is only a few percent. From the above considerations we obtain a range for
the $^{124}$Xe $2\nu$ECEC matrix element
based on our choice of parameters entering the ET.
Finally, we set the excitation energy to $\omega=478.3$~keV,
the average of the excitation energies of the lowest $2^+_1$
states in the corresponding even-even nuclei~\cite{coelloperez2017}.
This allows us to estimate the ET uncertainty associated to the
SSD approximation, Eq.~(\ref{SSD_relative}).

{\it Nuclear shell model.---}Next we perform large-scale shell model
calculations to obtain the nuclear matrix element
using the full expression Eq.~(\ref{M2EC_exact}).
We solve the many-body Schr\"odinger equation $H\left|\psi\right\rangle=E\left|\psi\right\rangle$
for $^{124}$Xe, $^{124}$Te, $^{124}$I, using a shell model Hamiltonian $H$
in the configuration space comprising the $0g_{7/2}$, $1d_{5/2}$, 
$1d_{3/2}$, $2s_{1/2}$, and $0h_{11/2}$ single-particle orbitals for neutrons 
and protons. To keep the dimensions of the shell model diagonalization tractable, 
especially for the largest calculation $^{124}$Xe, we need a truncated 
configuration space. In a first truncation scheme, similar to the one used 
in Ref.~\cite{klos2013}, we limit to two the number of nucleon excitations 
from the lower energy $0g_{7/2}$, $1d_{5/2}$ orbitals to the higher lying 
$1d_{3/2}$, $2s_{1/2}$, $0h_{11/2}$ orbitals. Second, we adopt a 
complementary truncation scheme that keeps a maximum of two neutron 
excitations but does not limit the proton excitations from lower to higher lying 
orbitals (a maximum of six nucleon excitations are permitted in $^{124}$Xe).
This keeps the $0g_{7/2}$ orbital fully occupied.
A third scheme with the $1d_{5/2}$ orbital fully occupied gives results 
within those of the other two truncations.

We use the shell model interaction GCN5082~\cite{menendez2009-1,caurier2010},
fitted to spectroscopic properties of nuclei in the mass region of $^{124}$Xe.
The shell model interaction has been tested against experimental data on Gamow-Teller
decays and charge-exchange transitions in this region, showing a good description 
of data with a renormalization, or ``quenching", of the $\bm \sigma \bm \tau$ 
operator $q=0.57$~\cite{caurier2012}. For the two-neutrino $\beta\beta$ decay 
of $^{128}$Te, $^{130}$Te, and $^{136}$Xe, however, this interaction fits data
best after a larger renormalization $q=0.48$~\cite{caurier2012}. An extreme
case is the very small $\beta\beta$ $^{136}$Xe matrix element, which is only
reproduced with $q=0.42$~\cite{caurier2012}. The renormalization of the spin-isospin
operator is needed to correct for the approximations made in the many-body 
calculation, such as unaccounted correlations beyond the configuration space 
or neglected two-body currents~\cite{engel2017,menendez2011}. A full understanding
of its origin would require an ab initio study that is currently possible only for
nuclei lighter than xenon~\cite{gazit2009,klos2017,ekstrom2014,pastore2017}.
Here we follow the strategy of previous shell model $\beta\beta$ decay
predictions~\cite{caurier1990,Poves1995} and include the above ``quenching"
factors phenomenologically to predict the half-life of $^{124}$Xe.

\begin{figure}[t]
\centering
\includegraphics[width=\columnwidth]{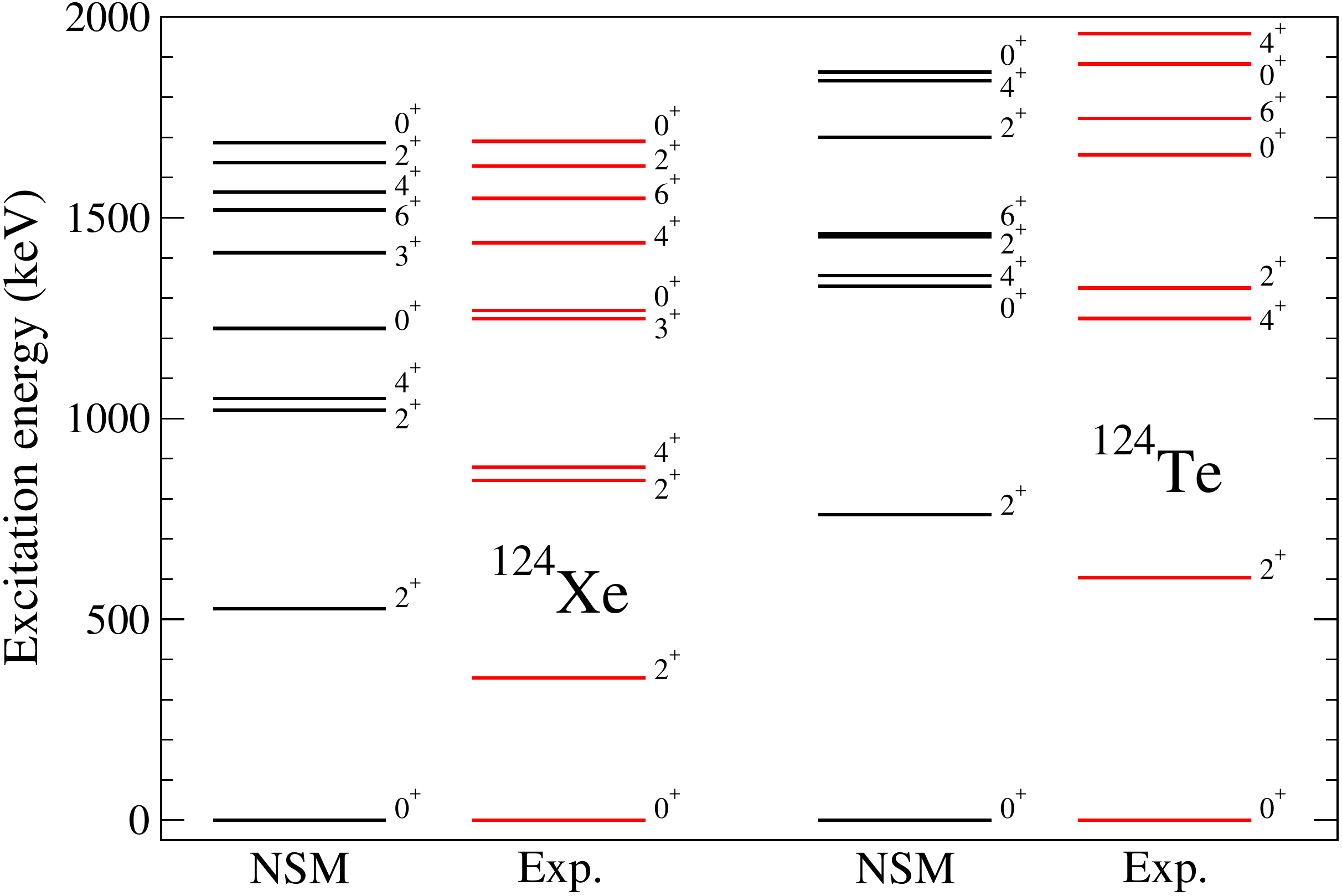}
\caption{$^{124}$Xe and $^{124}$Te excitation spectra obtained by the nuclear shell 
model (NSM) compared to experiment~\cite{nndc}.}
\label{spectra}
\end{figure}

The low-energy excitation spectra of the three isotopes are well
reproduced. Figure~\ref{spectra} compares the experimental and calculated 
spectra for $^{124}$Xe and $^{124}$Te, obtained with the first truncation 
scheme described above. The spectra corresponding to the second truncation 
scheme is of similar quality. When additional excitations to the higher lying 
orbitals are permitted, the first excited $0^+_2$ in $^{124}$Te is raised to 
1.6~MeV, in much better agreement with experiment. However, such extended
truncation yields too large dimensions for $^{124}$Xe, and cannot be used 
in our $2\nu$ECEC calculations. The spectra of the intermediate $^{124}$I 
is not well known besides the ground state and few tentative spin assignments.
The GCN5082 interaction reproduces correctly the spin and parity of the $2^-$
ground state, although with a lowest $1^+_1$ state at only about 10~keV, 
below any measured level. For the $2\nu$ECEC, as in the ET calculation,
we consider the lowest $1^+_1$ state at $105-170$~keV excitation energy.
All shell model calculations have been performed with the codes ANTOINE
and NATHAN~\cite{caurier1999,caurier2004}.

{\it Results and discussion.---}The calculated nuclear matrix elements
are common for the capture of K- or L-shell electrons.
However, the presented half-lives correspond to the $^{124}$Xe 
$2\nu{\rm EC EC}$ of two K-shell electrons, as this is the mode explored
in recent experiments~\cite{abe2016,aprile2017,abe2018,Gavrilyuk2018,Gavrilyuk2017}.

\addtolength{\tabcolsep}{2.5pt}
\begin{table}[t]
\caption{Nuclear matrix elements ($M^{2\nu{\rm ECEC}}$) and half-lives 
($T^{2\nu{\rm ECEC}}_{1/2}$) calculated with the ET and the nuclear shell model (NSM).
Results are given for the $^{124}$Xe $2\nu$ECEC of two K-shell electrons into the
$^{124}$Te ground $0^+_{\rm gs}$ and excited $0^+_2$ and $2^+_1$ states. 
The phase-space factors $G^{2\nu{\rm ECEC}}$ in y$^{-1}$ are from 
Refs.~\cite{kotila2013, pirinen2015}.}
\centering
\begin{tabular*}{\columnwidth}{c | c c c}
\hline\hline
$2\nu$ECEC
& $0^+_{\rm gs,i}\rightarrow 0^+_{\rm gs,f}$
& $0^+_{\rm gs,i}\rightarrow 0^+_2$
& $0^+_{\rm gs,i}\rightarrow 2^+_1$ \\
\hline
$G^{2\nu{\rm ECEC}}$ & $1.72\times10^{-20}$ & $1.67\times10^{-22}$ & $1.38\times10^{-23}$ \\
\hline
& \multicolumn{3}{c}{$M^{2\nu{\rm ECEC}}$} \\
ET & $0.011-0.041$ & $0.002-0.050$ & $(0.8-9.0)\!\times\!10^{-4}$ \\
NSM& $0.028-0.072$ & $0.005-0.010$ & $(1.1-2.3)\!\times\!10^{-4}$ \\ 
\hline
& \multicolumn{3}{c}{$T^{2\nu{\rm ECEC}}_{1/2}$} \\
& [$10^{22}$y] & [$10^{25}$y] & [$10^{30}$y] \\
ET &	$1.3-18$ & $0.092-57$ & $0.034-4.3$ \\
NSM & $0.43-2.9$ & $2.3-9.3$& $0.57-2.5$ \\
\hline\hline
\end{tabular*}
\label{results}
\end{table}

Table~\ref{results} summarizes our main results.
The ET predicts a smaller central value for the $^{124}$Xe $2\nu{\rm EC EC}$
matrix element than the NSM, even though both results are consistent
when taking uncertainties into account.
The ET uncertainty results from combining the uncertainty
associated to the SSD approximation, Eq.~(\ref{SSD_relative}),
with the range of the parameters used as input for the ET.
Both contributions are of similar size. For the NSM,
one part of the theoretical uncertainty is given
by the range of results obtained with different
truncation schemes.
The dominant part, however, is given
by the three ``quenching" values considered:
the average $q=0.57$ and $q=0.48$, corresponding to
the best description of Gamow-Teller transitions and
$\beta\beta$ decays, respectively, plus the additional
conservative $q=0.42$ needed in the $^{136}$Xe $\beta\beta$ decay.
The NSM ranges in Table~\ref{results}
cover the results obtained with the two truncations
and three ``quenching" values.

Table~\ref{results} also shows our predictions for the $2\nu$ECEC into
excited states of $^{124}$Te.
For both final $0^+_2$ and $2^+_1$ states,
the ET and NSM matrix elements are consistent,
even though the central values predicted by the shell model
are about one third of the ET ones.
The suppressed NSM matrix element to the final $0^+_2$ state
with respect to the transition to the ground state
is consistent with the results on neutrinoless $\beta\beta$ decay
in $^{128,130}$Te and $^{136}$Xe, using the same interaction~\cite{menendez2009-1}.
While the shell model uncertainties
are somewhat smaller than in the $2\nu$ECEC to the ground state,
the ET ones are much larger,
because of the limitations of the SSD approximation
when the energy denominator $D(1^+_1)$ is small~\cite{coelloperez2017}. 
The ET and NSM half-lives are in general shorter
than the QRPA ones
for the $0^+_2$ $2\nu$ECEC~\cite{suhonen2013,pirinen2015},
while for the $2^+_1$ $2\nu$ECEC the NSM and
QRPA~\cite{pirinen2015} predictions are very similar.
Transitions to excited states are extremely suppressed
because of the reduced $Q$-value and corresponding phase-space factor.
The $2\nu$ECEC to the final $2^+_1$ state,
which requires the capture of K- and L-shell electrons,
is further suppressed because of the small nuclear matrix element.

\begin{figure}[t]
\centering
\includegraphics[width=\columnwidth]{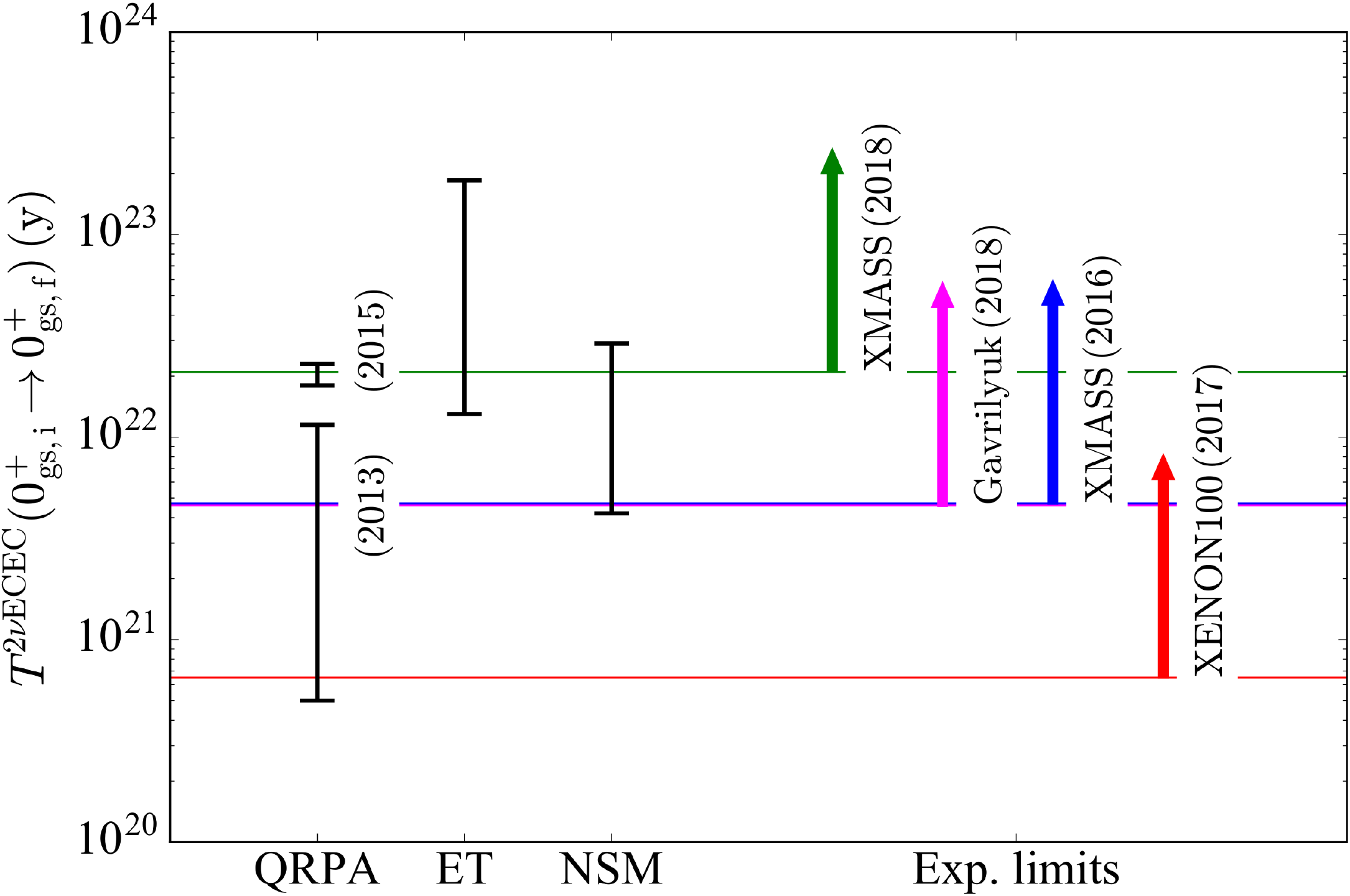}
\caption{$^{124}$Xe half-life for the $2\nu{\rm EC EC}$ of two K-shell electrons.
The black bars show the theoretical predictions from the effective theory (ET) and
the nuclear shell model (NSM), as well as most recent QRPA 
calculations~\cite{suhonen2013, pirinen2015}, in comparison to the horizontal
lines that indicate the experimental lower limits set by the XENON100~\cite{aprile2017}
(red) and XMASS~\cite{abe2016, abe2018} (blue, green) collaborations as
well as Gavrilyuk et al.~\cite{Gavrilyuk2018} (purple).}
\label{124Xe01}
\end{figure}

Figure~\ref{124Xe01} compares our theoretical predictions
for the $2\nu$ECEC on
$^{124}$Xe to the $^{124}$Te ground state
with the most advanced QRPA results from Refs.~\cite{suhonen2013, pirinen2015}
and the most recent experimental $2\nu$ECEC limits~\cite{abe2016,abe2018,aprile2017,Gavrilyuk2018}.
Theoretical half-lives are shown as black bars.
The predictions from the ET, NSM, and QRPA are consistent.
However, the ET shows a clear preference for longer half-lives than the NSM.
On the other hand, the QRPA spans much shorter half-lives
than those predicted by the ET or NSM.

Figure~\ref{124Xe01} shows that the theoretical predictions
are consistent with the lower half-life limits established by
the first results of the XMASS~\cite{abe2016} and
XENON100~\cite{aprile2017} collaborations
and with Ref.~\cite{Gavrilyuk2018},
shown as red, blue and purple horizontal lines in Fig.~\ref{124Xe01}, respectively.
However, the most recent limit established very recently by
XMASS~\cite{abe2018} (green line)
excludes most of our NSM results, but a part of the predicted range
remains permitted.
Note that since the shell model configuration space
had to be truncated, we could not obtain the exact nuclear matrix element
without ``quenching".
On the other hand, the ET half-life is almost fully
consistent with the current XMASS limit.
The ET central half-life is only about five times longer,
and the range predicted by the ET lies largely within
the sensitivity of ongoing experiments~\cite{aprile2017}.
The QRPA predictions are mostly excluded including error bars,
except the very recent results from Ref.~\cite{pirinen2015},
just at the border of the permitted region.
Most other older theoretical calculations are also
in tension with the XMASS limit~\cite{abe2018}.
Overall, our results suggest that the $^{124}$Xe $2\nu$ECEC
could very well be discovered in ongoing or upcoming experiments in the near future.
 
{\it Summary.---}We have calculated the nuclear matrix elements for the 
$2\nu$ECEC on $^{124}$Xe using an ET and the large-scale nuclear shell model,
two of the nuclear many-body approaches best suited to describe $\beta$ and 
EC transitions in heavy nuclei. The ET results are based on $\beta$ decay  
and EC on neighboring nuclei, while the shell model uses an interaction that 
describes well $\beta\beta$ decays of neighboring nuclei. The ET provides 
consistent theoretical uncertainties set by the order of the ET calculation,
while the shell model uncertainty is dominated by the range of ``quenching" 
considered for the $2\nu$ECEC operator. The ET predicts a half-life consistent 
and up to several times longer than current experimental limits, while the shell 
model prediction extends less beyond current limits. When all 
uncertainties are taken into account, the ET and NSM results are consistent,
as well as with the most advanced QRPA results.

Future directions include higher-order calculations in the ET to
reduce the uncertainties, and improved NSM studies with a better understanding 
of the ``quenching" of the operator, and limiting truncations in the configuration
space. Our findings suggest that the $^{124}$Xe $2\nu$ECEC has a good chance
to be discovered by ongoing or future experiments, so that these predictions
can be tested by upcoming analyses of ongoing experiments and can further
stimulate future searches.

We thank Alex Fieguth for useful discussions. This work was supported
in part by the Deutsche Forschungsgesellschaft under Grant SFB 1245, the
Japanese Society for the Promotion of Science through Grant 18K03639,
MEXT as Priority Issue on Post-K Computer (Elucidation of the
Fundamental Laws and Evolution of the Universe), and JICFus.

\end{document}